\documentclass[preprint, superscriptaddress, showpacs,preprintnumbers,amsmath,amssymb,prb]{revtex4}
\usepackage{graphicx}

\begin{document}

\thispagestyle{empty}

\title{Measuring the Casimir force gradient from
 graphene on
 a {\boldmath$\mbox{SiO}_2$} substrate}
\author{
A.~A.~Banishev}
\affiliation{Department of Physics and Astronomy, University of California, Riverside, California 92521, USA}
\author{
H.~Wen}
\affiliation{Department of Physics and Astronomy, University of California, Riverside, California 92521, USA}
\author{
J.~Xu}
\affiliation{Department of Physics and Astronomy, University of California, Riverside, California 92521, USA}
\author{
R.~K.~Kawakami}
\affiliation{Department of Physics and Astronomy, University of California, Riverside, California 92521, USA}
\author{
G.~L.~Klimchitskaya}
\affiliation{Central Astronomical Observatory at Pulkovo of the Russian Academy of Sciences, St.Petersburg, 196140, Russia}
\author{
V.~M.~Mostepanenko}
\affiliation{Central Astronomical Observatory at Pulkovo of the Russian Academy of Sciences, St.Petersburg, 196140, Russia}
\author{
 U.~Mohideen}
\affiliation{Department of Physics and Astronomy, University of California, Riverside, California 92521, USA}

\begin{abstract}
The gradient of the Casimir force between a Si-SiO${}_2$-graphene
substrate and an Au-coated sphere is measured by means of a
dynamic atomic force microscope operated in the frequency shift
technique. It is shown that the presence of graphene leads to up
to 9\% increase in the force gradient at the shortest separation
considered. This is in qualitative agreement with the predictions
of Lifshitz theory using the dielectric permittivities of Si
and SiO${}_2$ and the Dirac model of graphene.
\end{abstract}

\maketitle

\section{Introduction}

In the last few years graphene has attracted considerable
attention
as a material of much promise for nanotechnology due to
its unique mechanical, electrical and optical
properties.\cite{1,2} Noting that at short separations
between test bodies the fluctuation-induced
dispersion interactions, such as the van der Waals and
Casimir forces, become dominant,\cite{2a} it is important to
investigate them in the presence of a graphene sheet.
In this connection much theoretical  work has been
done on the calculation of dispersion forces between two
graphene sheets,\cite{3,3a,4,5,6,7a} a graphene sheet and
a metallic, dielectric or semiconductor
plate,\cite{4,5,6,7,8,9,10,11,12} a graphene sheet and an
atom or a molecule\cite{13,14,15,16} etc.
The calculations were performed using phenomenological
density-functional methods,\cite{17,18,19,20}
second order perturbation theory,\cite{21} and the
Lifshitz theory with some specific form for the
reflection coefficients of electromagnetic oscillations
on graphene.\cite{3,5,8,9} However, in spite of the
impressive progress in measurements of the Casimir
force in configurations with metallic, dielectric and
semiconductor test bodies (see reviews in
Refs.~\cite{22,23,24,25} and more recent experiments
\cite{26,27,28}),  there is yet
no previous measurement of dispersion forces acting on
graphene.

In the present paper we report measurements of the
gradient of the Casimir force acting between a graphene
sheet deposited on a SiO${}_2$ film covering a Si plate
and an Au-coated sphere. Our measurements are performed
by means of dynamic atomic force microscope (AFM)
operated in the frequency-shift technique described in
detail in Refs.~\cite{27,28}. We demonstrate significant
increase in the gradient of the Casimir force in
comparison with that between a Si plate covered with a
SiO${}_2$ film and an Au-coated sphere, i.e., in the
absence of graphene sheet. At short separations this
increase is up to a factor 4-5 larger than the total
experimental error in the measurement of the force gradient
determined at a 67\% confidence level.
We also compare the experimental results with an
approximate theory where the gradients of the Casimir
force between a Si-SiO${}_2$ system and Au-coated sphere
and between a graphene described by the Dirac model
and the same sphere are computed independently of one
another using the Lifshitz theory and then are added.
Some excess of the theoretical force gradient over
the experimental one is attributed to the screening
of the Si-SiO${}_2$ surface by a graphene sheet.

The paper is organized as follows. In Sec.~II we briefly
describe the detection system, the measurement scheem and
the sample preparation. Section III contains the measurement
results and their comparison with  theory. Section IV
contains our conclusions.

\section{Experimental setup}

The detection system used in our measurements consists
of an AFM cantilever with attached hollow glass
microsphere coated with Au, piezoelectric actuators,
fiber interferometers, light source, and phase locked
loop (PLL). The thickness of the Au coating and the
radius of the coated sphere were measured to be 280\,nm
and $54.10\pm 0.09\,\mu$m using an AFM and a scanning
electron microscope, respectively.
A turbo-pump, oil-free dry scroll
mechanical-pump
and ion-pump were used to achieve high vacuum down
to $10^{-9}\,$Torr (see Refs.~\cite{27,28} for detail of
the setup).

In the dynamic measurement scheme the total force
$F_{\rm tot}(a)=F_{\rm el}(a)+F(a)$ acting on the
sphere [where $F_{\rm el}(a)$ and $F(a)$ are the
electric and Casimir force, respectively, and $a$
is the separation distance between the sphere and
graphene] modifies the resonant natural frequency of the
oscillator. The change in the frequency
$\Delta\omega=\omega_r-\omega_0$,
where $\omega_r$ and $\omega_0$ are the resonance
frequencies in the presence and in the absence of
external force $F_{\rm tot}(a)$, respectively, was
recorded by the PLL. This was done at every 0.14\,nm
while the plate was moved towards the
grounded sphere starting at the maximum separation.
This was repeated with
one of 10 different voltages $V_i$ in
the range from --38.5 to 58.4\,mV for the first graphene
sample and from --5.2 to 97.4\,mV for the second graphene
sample applied to the graphene sheet
while the sphere remained grounded.
The application of voltages and respective measurements
were repeated for two times resulting in 20 sets of
$\Delta\omega$ as a function of separation for each
graphene sample.

Large area graphene used in our experiment was obtained
through a two-step Chemical Vapor Deposition (CVD) process
described.\cite{29} In this process $25\,\mu$m
thick polycrystalline copper foil (99.8\% purity) was
cleaned by diluted HCl solution followed by deionized
water rinse. Then the copper foil was placed into
$\sim 5\,\mbox{cm}\times 3\,$cm copper bag which
had undergone the same clean process as above.
The copper bag was loaded into a ceramic tube furnace for
the CVD process. First the copper bag was annealed at
$1000^{\circ}$C under continuous Ar/H${}_2$
(69\,sccm/10\,sccm) flow. Graphene was grown on the copper foil
by introducing methane/hydrogen gas  of 1.3\,sccm/4\,sccm
for one hour and
35\,sccm/4\,sccm for another hour.
Then the furnace was cooled down to room temperature
under a continuous flow of  Ar/H${}_2$ (69\,sccm/10\,sccm).
Finally, the grown graphene was transferred from the copper foil
to 300\,nm SiO${}_2$ layer on a B-doped Si layer of $500\,\mu$m
thickness on the bottom by using poly-metil methacrylate
(PMMA) as the graphene support layer and ammonium persulfate
solution as the copper etchant.
We have examined the quality of the graphene layer
through Raman spectroscopy \cite{30a,30b} and quantum Hall effect
measurements,\cite{30c,30d} which show single layer graphene
characteristics.
Measurements of 2D-mobility for a large area graphene onto
SiO${}_2$ substrates performed in our laboratory demonstrate
mobility above $3000\,\mbox{cm}^2/\mbox{Vs}$.
A roughly estimate for the concentration of impurities would be
$1.2\times 10^{10}\,\mbox{cm}^{-2}$, if we consider that each
impurity adsorbs one electron.

The gradients of the total and Casimir forces were found
from the measured frequency shifts using electrostatic
calibration. To perform this calibration of the setup, we
used the expression for the electric force in
sphere-plate geometry \cite{23}
\begin{equation}
F_{\rm el}(a)=X(a,R)(V_i-V_0)^2.
\label{eq1}
\end{equation}
\noindent
Here $X(a,R)$ is a known function and $V_0$ is the
residual potential difference between a sphere surface and
a graphene sheet which is nonzero even when both surfaces
are grounded. In the linear regime which is realized in our setup
\cite{27} the gradient of the Casimir force is given by
\begin{equation}
F^{\prime}(a)\equiv\frac{\partial F(a)}{\partial a} =
-\frac{1}{C}\Delta\omega -\frac{\partial X(a,R)}{\partial a}
(V_i-V_0)^2,
\label{eq2}
\end{equation}
\noindent
where $C=\omega_0/(2k)$ and $k$ is the spring constant of the
cantilever. Note that the absolute separations between the
zero level of the roughness on the sphere and graphene are
found from $a=z_{\rm piezo}+z_0$, where $z_{\rm piezo}$ is
the plate movement due to the piezoelectric actuator and
$z_0$ is the closest approach between the Au sphere and
graphene (in dynamic experiments the latter is much larger
than the separation on contact of the two surfaces).

\begin{figure}[b]
\vspace*{-15cm}
\centerline{\hspace*{3cm}
\includegraphics{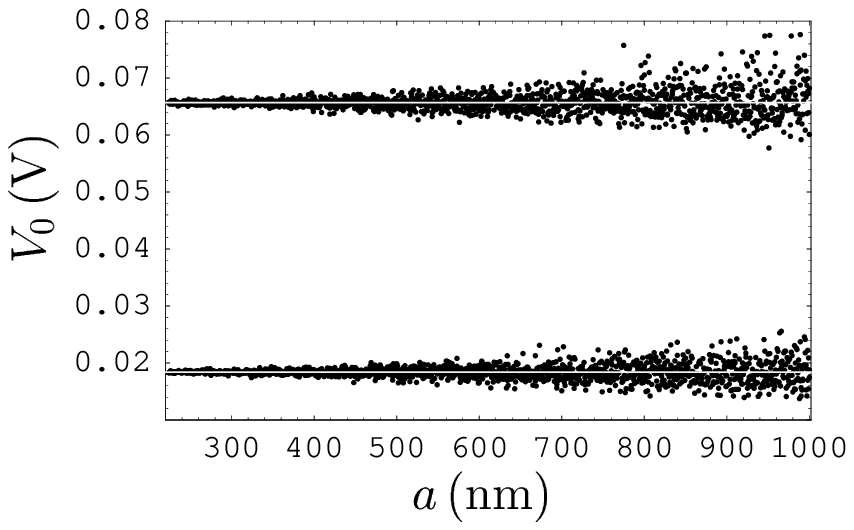}
}
\vspace*{-7cm}
\caption{\label{fg1}The residual potential difference between an
Au-coated sphere and the first (lower dots) and second (upper dots)
graphene sheets on a Si-SiO${}_2$ substrate as a function of
separation. The mean values of $V_0$ are shown by the gray lines.
}
\end{figure}

{}From the position of a maximum in the parabolic dependence
of $\Delta\omega$ on $V_i$ in Eq.~(\ref{eq2}), one can determine
$V_0$ with the help of a $\chi^2$-fitting procedure.
{}From the curvature of the parabola with the help of the same
fit it is possible to determine $z_0$ and $C$. This was done
at different separations for the two graphene samples used
in our experiment. In Fig.~\ref{fg1} we present the values
of $V_0$ as a function of separation determined from the fit
for the first and second graphene samples (the lower and upper
sets of dots, respectively). The obtained values were
corrected for mechanical drift of the frequency-shift signal,
as discussed in Ref.~\cite{27}. As can be seen from
Fig.~\ref{fg1},
the resulting $V_0$ do not depend on separation.
To check this observation, we have performed the best fit of
$V_0$ to the straight lines shown in Fig.~\ref{fg1} leaving
their slopes as free parameters. It was found that the slopes are
$-4.96\times 10^{-6}\,$mV/nm and $6.2\times 10^{-4}\,$mV/nm
for the first and second samples, respectively, i.e., the
independence of $V_0$ on $a$ was confirmed to a high
accuracy. This finally leads to the mean values
$V_0=18.4\pm 0.9\,$mV and $V_0=65.7\pm 0.9\,$mV for the
first and second samples, respectively, where errors are
determined at a 67\% confidence level.
Note that different graphene sheets may lead to different
$V_0$ due to occasional impurities.
The possible impurities could be organic, H${}_2$, O${}_2$, N${}_2$
 and H${}_2$O. All these may
become dopants of graphene and change its work
function.
Next the quantities $z_0$ and $C$ were determined from the fit at
different separations and found to be separation-independent.
For the first and second samples the mean values are equal to
$z_0=222.5\pm 0.4\,$nm, $C=58.7\pm 0.17\,$kHz\,m/N and
$z_0=222.2\pm 0.4\,$nm, $C=58.9\pm 0.17\,$kHz\,m/N,
respectively.
{}From the measured resonant frequency
we have confirmed that  the obtained value of $C$ results in
the spring constant $k$ consistent with the
estimated value provided by the cantilever fabricator.

\section{Measurement results and comparison with theory}

\begin{figure}[b]
\vspace*{-5cm}
\centerline{\hspace*{3cm}
\includegraphics{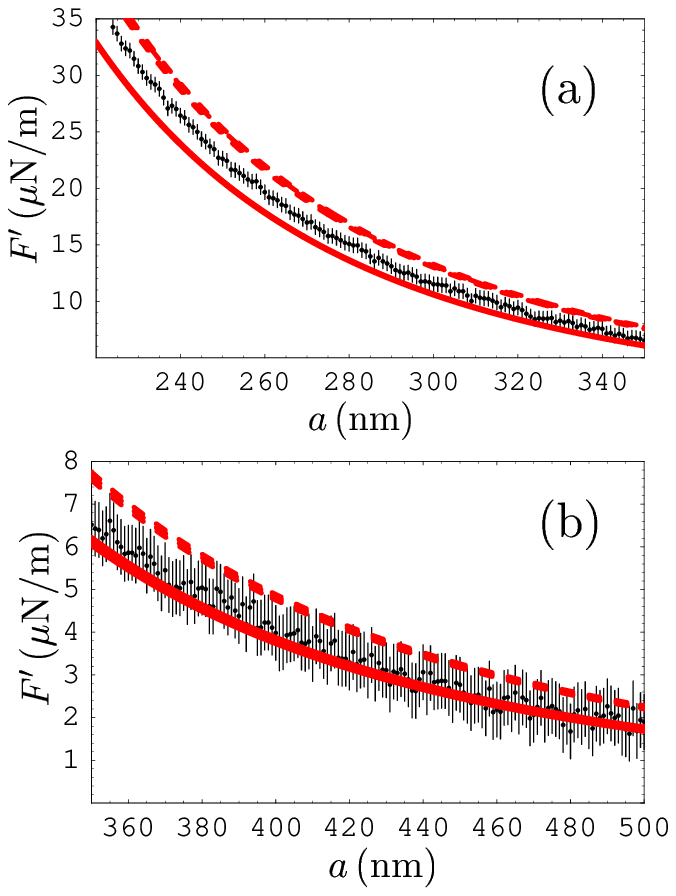}
}
\vspace*{-12cm}
\caption{\label{fg2}(Color online) The experimental data for
the gradient of the Casimir force $F^{\prime}$
at (a) short and (b) long separations
are shown as crosses plotted at a 67\% confidence
level (measured with the applied compensating voltage for
the first sample).
The theoretical $F^{\prime}$ between an Au-coated
sphere and a Si-SiO${}_2$ substrate calculated using
the Lifshitz theory and between an Au-coated sphere and
graphene deposited on this substrate calculated using
an additive approach
are shown as the solid and dashed bands, respectively.
}
\end{figure}

For each graphene sample the gradients of the Casimir force
$F^{\prime}(a)$ as a function of $a$ were obtained from the
measured $\Delta\omega$ in two ways:
by applying 10 different voltages $V_i$ with subsequent
subtraction of the electric forces (2 repetitions) and
by applying the compensating voltage $V_i=V_0$ (22 repetitions).
In these cases 20 and 22 force-distance relations were obtained,
the mean force gradients were computed and their total experimental
errors were determined at a 67\% confidence level as a
combination of random and systematic errors (see
Ref.~\cite{27} for details).
In Fig.~\ref{fg2}(a,b) the mean gradients of the Casimir force
and their errors measured for the first sample with applied
compensating voltage are shown as crosses with a step of 1\,nm.
Table ~1 presents the values of mean $F^{\prime}(a)$ at
several separations measured in the two different ways for the first
(columns a,\,b) and second (columns c,\,d) samples.
As can be seen in Table~1, the measurement results for the two
graphene samples obtained in two different ways are in very good
mutual agreement.

Now we compare the experimental results with theoretical
predictions. At the moment there is no theory allowing  rigorous
calculation of the Casimir force between a graphene deposited on
a Si-SiO${}_2$ substrate and an Au sphere. The problem is that
Si and SiO${}_2$ layers are described by their dielectric
permittivities and the reflection properties of
graphene in the Dirac model are described by the polarization
tensor.
This does not allow direct
application of the Lifshitz theory for layered
structures \cite{23,30}. Because of this, here we restrict
ourselves to the approximate approach, where the contributions of
Si-SiO${}_2$ substrate  and graphene sheet to the Casimir interaction
with an Au sphere are computed separately using the Lifshitz
theory and are then added together.
In the framework of the proximity force approximation (PFA),
the Lifshitz formula for the gradient of the Casimir force between
an Au sphere and any planar structure takes the form
\begin{equation}
F^{\prime}(a)=2k_BTR\sum_{l=0}^{\infty}{\vphantom{\sum}}^{\prime}
\!\!\int_{0}^{\infty}\!\!\!\!
q_lk_{\bot}dk_{\bot}\sum_{\alpha}
\frac{r_{\alpha}^{(1)}r_{\alpha}^{(2)}}{e^{2q_la}-r_{\alpha}^{(1)}r_{\alpha}^{(2)}}.
\label{eq3}
\end{equation}
\noindent
Here $k_B$ is the Boltzmann constant, $T=300\,$K is the
laboratory temperature, $k_{\bot}$ is the projection of
the wave vector on a planar structure,
$q_l^2=k_{\bot}^2+\xi_l^2/c^2$, and $\xi_l=2\pi k_BTl/\hbar$
with $l=0,\,1,\,2,\,\ldots$ are the Matsubara frequencies.
The prime near the summation sign multiplies the term with
$l=0$ by 1/2, and $\alpha={\rm TM,TE}$ denotes the transverse
magnetic and transverse electric polarizations of the
electromagnetic field.
Note that an error arising from the application of PFA was
recently found \cite{31,32,33} using the exact theory for
the sphere-plate geometry and was shown to be less than
$a/R$, i.e., of about 0.5\% in our experiment.

The quantity $r_{\alpha}^{(1)}=r_{\alpha}^{(1)}(i\xi_l,k_{\bot})$
in Eq.~(\ref{eq3}) is the standard Fresnel reflection coefficient
for an Au surface calculated at the imaginary frequencies (an Au layer
can be considered as a semispace). It is expressed in terms of
the dielectric permittivity
$\varepsilon^{\,\rm Au}(i\xi_l)$ using the tabulated optical
data for Au \cite{34} extrapolated to zero frequency either by the
Drude or by the plasma models.\cite{22,23}

Unlike the case when a graphene layer is present, the Casimir
interaction
of the Si-SiO${}_2$ substrate with an Au sphere is described by
the well tested fundamental Lifshitz theory.
Here the quantity
$r_{\alpha}^{(2)}=r_{\alpha}^{(2)}(i\xi_l,k_{\bot})$
has the meaning of the reflection coefficient on the two-layer
(Si-SiO${}_2$) structure \cite{23,30,35} where Si can be
considered as a semispace.
It is expressed in terms of
$\varepsilon^{\,\rm Si}(i\xi_l)$ and
$\varepsilon^{\,{\rm SiO}_2}(i\xi_l)$.
In our computations we used $\varepsilon^{\,\rm Si}(i\xi_l)$
obtained \cite{36} from the optical data \cite{37} for Si
extrapolated to zero frequency either by the Drude or by the
plasma models (Si plate used has the resistivity between
0.001 and $0.005\,\Omega\,$cm which corresponds to a plasma
frequency between $5\times 10^{14}$ and $11\times 10^{14}\,$rad/s
and the relaxation parameter
$\gamma\approx1.1\times 10^{14}\,$rad/s).
A sufficiently accurate expression for
$\varepsilon^{\,{\rm SiO}_2}(i\xi_l)$ from Ref.~\cite{38}
was used in the computations.
The r.m.s. roughness on the surfaces of sphere and graphene
was measured by means of AFM and found to be equal to 1.6\,nm
and 1.5\,nm, respectively. It was taken into account using
the multiplicative approach,\cite{22,23} and its maximum
contribution to the force gradient is equal to only 0.1\% at
the shortest separation.

The computational results for $F^{\prime}(a)$ between a
Si-SiO${}_2$ substrate and an Au sphere are shown by the
solid band in Fig.~\ref{fg2}. The width of the band indicates
the uncertainty in the value of $\omega_p$ and a difference between
the predictions of the Drude and plasma model approaches to the
description of Au and Si which is small in this
experiment. The latter is illustrated in columns e and f of
Table~1. Figure~\ref{fg2} and Table~1 indicate conclusively
that within the separation region from 224 to 320\,nm the
measured gradients of the Casimir force are larger than that
for a Si-SiO${}_2$ substrate interacting with an Au sphere.
This demonstrates the influence of the graphene sheet on the
Casimir force.

\begin{figure}[b]
\vspace*{-15cm}
\centerline{\hspace*{3cm}
\includegraphics{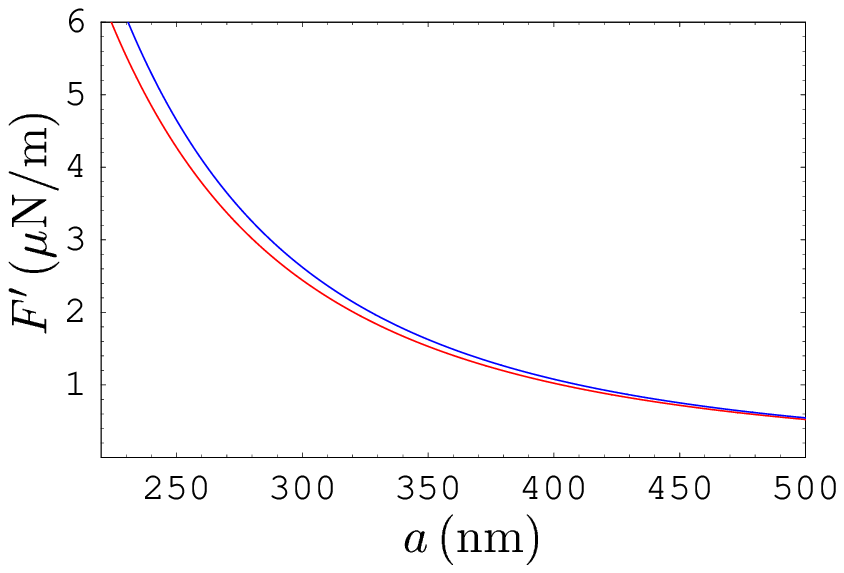}
}
\vspace*{-7cm}
\caption{\label{fg3}(Color online)
The gradient of the Casimir force
between an Au-coated sphere and a suspended graphene sheet
calculated using the Dirac model with the mass gap parameter
equal to 0.1\,eV (lower line) and 0 (upper line) as a function
of separation.
}
\end{figure}

The reflection coefficients for a suspended
graphene described by the Dirac model are represented in
the form \cite{9,11,16}
\begin{eqnarray}
&&
r_{\rm TM}^{(2)}=
\frac{q_l\Pi_{00}}{q_l\Pi_{00}+2\hbar k_{\bot}^2},
\label{eq4} \\
&&
r_{\rm TE}^{(2)}=
-\frac{k_{\bot}^2\Pi_{\rm tr}-
q_l^2\Pi_{00}}{k_{\bot}^2(\Pi_{\rm tr}+2\hbar q_l)-q_l^2\Pi_{00}},
\nonumber
\end{eqnarray}
\noindent
where $\Pi_{mn}$ are the components of the polarization tensor in
3-dimensional space-time and the trace stands for the sum of spatial
components. The computational results for the gradient of the
Casimir force between the suspended graphene with the mass gap
parameter $\Delta=0$ and $\Delta=0.1\,$eV and an Au sphere as
a function of $a$ are shown in Fig.~\ref{fg3} by the upper and
lower lines, respectively (here the results do not depend on
whether the Drude or the plasma model approach for Au is
used \cite{11}). In Fig.~\ref{fg2} the dashed band shows the
sum of the force gradients between  a Si-SiO${}_2$ substrate
and an Au sphere and between graphene and the same sphere.
The width of the band takes into account the respective width for
a substrate interacting with a sphere
and also differences in predictions of the Dirac model
of graphene with the mass gap parameter varying from 0 to
0.1\,eV. It can be seen in Fig.~\ref{fg2} that the
used approximate
approach overestimates the measured force gradient, as it should,
keeping in mind that it does not take into account the
screening of the SiO${}_2$ surface by the graphene layer.
Thus our results also illustrate nonadditivity of the van der Waals
and Casimir interactions in multilayer structures.\cite{40}
Note that at short separations our approximate approach
(dashed line in Fig.~\ref{fg2}) is in better agreement with the
data than the approach which disregards the graphene layer
(solid line in Fig.~\ref{fg2}).
Thus, at $a=224\,$nm the relative difference between the
prediction of the approach disregarding graphene and the
measured force gradient is equal to --10.1\% of the measurement
result and between the prediction of our approximate approach
taking graphene into account and the same force gradient is
equal to 7.1\%. It is quite natural, however, that at large
separations the influence of the graphene layer is overestimated
by our approximate approach.

\section{Conclusions}

To conclude, we have demonstrated the influence of a graphene
layer on the Casimir force between a Si-SiO${}_2$ substrate
and an Au sphere. At the shortest separation measured the
relative excess in the force gradient due to the presence of
graphene deposited on a substrate reaches 9\% and decreases
with increasing separation.
Our experimental results are found to be in qualitative
agreement with an approximate theoretical
approach describing the reflection coefficients on
graphene via the polarization tensor in 3-dimensional
space-time, whereas the layers of the substrate are described by
means of the dielectric permittivity. The standard Lifshitz theory
for layered structures is not applicable to such cases.
A more exact theoretical description than the one used in this
work remains a challenge to theory.
The present work will serve as a motivation in this direction.
The Casimir interaction of graphene should be taken into
account in future applications of carbon nanostructures in
nanotechnology.

\section*{Acknowledgments}
This work was supported by the DOE Grant
No.\ DEF010204ER46131 (A.B., G.L.K., V.M.M., U.M.),
by the NRI-NSF Grant No.\ NEB--1124601 (graphene film,
H.W., R.K.K.)
and by the
NSF Grant No.\ PHY0970161 (equipment, G.L.K., V.M.M., U.M.).



\begingroup
\squeezetable
\begin{table}
\vspace*{2cm}
\caption{The mean values of the gradient of the Casimir force
together with their total experimental errors
at different separations (first column) measured in this work with
applied compensating voltage (column a) and with different applied
voltages (column b) for the first graphene sample (columns c and
d, respectively,
for the second graphene sample). Columns~e and f contain theoretical
values for the gradients of the Casimir force between the Au sphere
and Si-SiO${}_2$ substrate calculated when Au and Si are described
by the plasma and Drude model approaches, respectively.
}
\begin{ruledtabular}
\begin{tabular}{crrrrrr}
&\multicolumn{6}{c}{Gradients of the Casimir force
$F^{\prime}\,(\mu\mbox{N/m})$}
\\[1mm]
\cline{2-7}
$a\,$(nm)&a{\ \ \ \ }{\ \ \ \ }&b{\ \ \ \ }{\ \ \ \ }&
c{\ \ \ \ }{\ \ \ \ }&d{\ \ \ \ }{\ \ \ \ }&
e{\ \ \ \ }&f{\ \ \ \ }\\
\hline
224&$34.27\pm 0.64$&$33.58\pm 0.65$&$34.12\pm 0.64$&$33.76\pm 0.65$&
30.90&30.70\\
250&$22.62\pm 0.64$&$22.27\pm 0.64$&$22.72\pm 0.64$&$22.42\pm 0.64$&
20.67&20.51\\
300&$11.50\pm 0.64$&$11.19\pm 0.64$&$11.65\pm 0.64$&$11.53\pm 0.64$&
10.66&10.54\\
350&$6.52\pm 0.64$&$6.28\pm 0.64$&$6.30\pm 0.64$&$6.60\pm 0.64$&
6.12&6.03\\
400&$3.98\pm 0.64$&$3.67\pm 0.64$&$3.99\pm 0.64$&$3.70\pm 0.64$&
3.81&3.73\\
500&$1.90\pm 0.64$&$1.76\pm 0.64$&$1.78\pm 0.64$&$1.60\pm 0.64$&
1.73&1.68
\end{tabular}
\end{ruledtabular}
\end{table}
\endgroup
\end{document}